%% file: Letter_MIMO_Switch_04152.tex
\documentclass[12pt] {IEEEtran}
\onecolumn
\usepackage{setspace}
\input{pream_bm}

\usepackage{tikz}
\usetikzlibrary{arrows}
\tikzstyle{block}=[draw opacity=0.7,line width=1.4cm]

\usepackage{epsfig,latexsym,amssymb,amsmath,graphics}
\usepackage{graphicx}

\begin{document}
\IEEEoverridecommandlockouts
\title{Wireless MIMO Switching}

%\author{\IEEEauthorblockN{Fanggang Wang and \ Soung~C.~Liew}
%\IEEEauthorblockA{Institute of Network Coding, CUHK, Hong Kong, China.\\
%Email: fgwang@inc.cuhk.edu.hk,\ soung@ie.cuhk.edu.hk} }

\author{Fanggang~Wang and Soung Chang~Liew\\
%Institute of Network Coding\\The Chinese University of Hong Kong
\thanks{Corresponding author: Fanggang Wang, fgwang@inc.cuhk.edu.hk}}

\maketitle

\begin{abstract} \label{abs}
%\boldmath
In a generic switching problem, a switching pattern consists of a one-to-one mapping from a set of inputs to a set of outputs (i.e., a permutation). We propose and investigate a wireless switching framework in which a multi-antenna relay is responsible for switching traffic among a set of $N$ stations. We refer to such a relay as a MIMO switch. With beamforming and linear detection, the MIMO switch controls which stations are connected to which stations.   Each beamforming matrix realizes a permutation pattern among the stations. We refer to the corresponding permutation matrix as a switch matrix. By scheduling a set of different switch matrices, full connectivity among the stations can be established. In this paper, we focus on ``fair switching'' in which equal amounts of traffic are to be delivered for all $N(N-1)$ ordered pairs of stations.  In particular, we investigate how the system throughput can be maximized. In general, for large $N$ the number of possible switch matrices (i.e., permutations) is huge, making the scheduling problem combinatorially challenging. We show that for $N=4$ and $5$, only a subset of $N-1$ switch matrices need to be considered in the scheduling problem to achieve good throughput. We conjecture that this will be the case for large $N$ as well. This conjecture, if valid, implies that for practical purposes, fair-switching scheduling is not an intractable problem.
\end{abstract}


% \fi
\IEEEpeerreviewmaketitle

\section{Introduction}
Relaying in wireless networks plays a key role in various communication applications \cite{cov79}. The use of relays can extend coverage as well as improve energy efficiency \cite{den09}. In this paper, we study a set-up in which $N$ stations communicate with each other via a multi-antenna relay. With beamforming, the relay controls which stations are connected to which other stations. Each beamforming matrix realizes a permutation among the stations represented by a switch matrix. By scheduling a set of different switch matrices, full connectivity among the stations can be established.

%Network coding offers a new paradigm for networking and communications \cite{Ahl00} \cite{Li03} \cite{Yeung06}. The first attempt of applying network coding in wireless communication is in two-way relay channel. In \cite{Wu04informationexchange}, it was shown that a straightforward application of network coding can improve throughput by 33\% over the traditional store-and-forward  scheme. Ref. \cite{Zhang06physical-layernetwork} proposed physical-layer network coding (PNC) that exploits superposition of electromagnetic waves , boosting the throughput improvement to 100\%. In \cite{Katti07embracingwireless}, a simple implementation of PNC called analog network coding (ANC) was proposed.

Prior work that investigated $N$ stations exchanging data via a relay includes \cite{den09, moh09, Cui08}, and \cite{Gao09}. Ref. \cite{den09} studied ``pairwise data exchange'', in which stations form pairs, and two stations in a pair exchange data with each other only. Specifically for pairwise data exchange, if station $i$ transmits to station $j$, then station $j$ transmits to station $i$ as well. In \cite{den09}, MIMO relays with different forwarding strategies were considered. Ref. \cite{moh09} also studied pairwise data exchange, but the relay adopts the decode-and-forward strategy only. The diversity-multiplexing tradeoffs under reciprocal and non-reciprocal channels were analyzed. Both \cite{den09} and \cite{moh09} studied the case in which a station communicates with one other station only. In a general setting, a station could have data for more than one station. In this paper, we focus on a uniform traffic setting in which the amounts of traffic from station $i$ to station $j$ are the same for all $i,j\in \{1, \cdots, N\}$, $i \neq j$. ``Fair switching'' is used to meet the uniform traffic requirement. Specifically, fair switching is realized by scheduling a set of switch matrices. To the best of our knowledge, the framework of fair switching has not been considered in the existing literature.

Refs. \cite{Cui08} and \cite{Gao09} investigated the case of full data exchange, in which all stations want to broadcast their data to all other stations\footnote{Note that full data exchange is also discussed in \cite{den09}. But they consider a single-antenna relay.}. Data transmissions in \cite{Cui08} and \cite{Gao09} can be summarized as follows: in the first slot, all stations transmit to the relay simultaneously; the first slot is followed by multiple slots for downlink transmissions; in each downlink slot, the relay multiplies the signal received in the first time slot by a different beamforming matrix, such that at the end of all downlink slots, all stations receive the broadcast data from all other stations. By contrast, the framework investigated in this paper is more general in that it can accommodate the pure unicast case, the mixed unicast-multicast case, as well as the pure broadcast case as in \cite{Cui08} and \cite{Gao09}. In particular, a station $i$ can have $M_i$ data streams, and each station $j \neq i$ is a target receiver of one of the $M_i$ streams.

In our framework, the MIMO relay serves as a general switch that switches traffic among the stations. We use beamforming at the relay and linear detection to realize different connectivity patterns among the stations.
Each beamforming matrix realizes a permutation among the stations represented by a switch matrix. By scheduling a set of switch matrices, the MIMO switching system can realize any general transmission pattern (unicast, multicast, broadcast, or a mixture of them) among the stations.

Before delving into technical details, we provide a simple example to illustrate the scenario of interest to us here.  Consider a network with three stations, $1$, $2$, and $3$. The traffic flows among them are shown in Fig.~\ref{3node_demand}: station $1$ wants to transmit ``$a$'' to both stations $2$ and $3$; station $2$ wants to transmit ``$b$'' and ``$c$'' to stations $1$ and $3$, respectively; station $3$ wants to transmit ``$d$'' and ``$e$'' to stations $1$ and $2$, respectively. Pairwise data exchange as in \cite{den09} and \cite{moh09} is not effective in this case because when the number of stations is odd, one station will always be left out when forming pairs.  That is, when the number of stations is odd, the connectivity pattern realized by a switch/permutation matrix does not correspond to pairwise communication. Full data exchange is not appropriate either, since in our example, station $2$ (as well as station $3$) transmits different data to the other two stations. Under our framework, the traffic flows among stations can be met as shown in Fig.~\ref{3node_realization}. In the first slot, station $1$ transmits ``$a$'' to station $3$; station $2$ transmits ``$b$'' to station $1$; station $3$ transmits ``$e$'' to station $2$. In the second slot, station $1$ transmits ``$a$'' to station $2$; station $2$ transmits ``$c$'' to station $3$; station $3$ transmits ``$d$'' to station $1$.
%Regarding a fairness issue in the framework for practical application, the performance study in this paper considers the uniform amount of traffic pattern in which the traffic from station $i$ to station $j$ is the same for all $i$ and $j$, $i\neq j$. Note that this is not the full data exchange case in \cite{Cui08, Gao09} because a station transmits different data to different stations rather than broadcast the same data to them. To meet the demands of the uniform amount of traffic pattern, ``fair switching'' is required so that equal amounts of traffic are switched between all $N(N-1)$ ordered pairs of stations.
In Section III.C, we will present the details on how to realize the switch matrices.
To limit the scope, this paper focuses on the use of amplify-and-forward relaying and zero forcing (ZF) to establish the permutations among stations.
%However, we do generalize the ZF method to one that exploits physical-layer network coding \cite{Ahl00}, \cite{Li03}, \cite{Zhang06physical-layernetwork}, \cite{Katti07embracingwireless} for performance improvement.

%In this paper we propose a general framework for relay network with multiple stations, and call it \emph{wireless MIMO switching network}. It can realize any transmission pattern by designing a derangement matrix, which is a special permutation without diagonal elements. With the derangement transmission pattern, thus we propose a linear zero-forcing detection algorithm for the framework. In order to guarantee each station sends the same amount of data to each of the other stations in a complete transmission process, a fair switching is proposed, and one complete fair transmission consists of multiple rounds of switching. There are different ways of complete transmission, and they are evaluated in this paper to choose the best one. Furthermore, we relax the zero-forcing requirement for the diagonal elements of the derangement, and generalize the framework into a network coding-based scheme, namely \emph{MIMO switching with network coding}. Then we propose a naive method to solve the problem as an illumination of of the topic and leave a large room on how to optimize switched network coding to improve the performance of the fundamental wireless MIMO switching network.

The rest of the paper is organized as follows: Section II describes the framework of wireless MIMO switching and introduces the ZF relaying method for establishing permutations among stations. The fair switching framework is presented in Section III. Section IV discusses our simulation results. Section V concludes this paper.

\section{System Description} \label{sec.SystemDes}

\subsection{System Model}
Consider $N$ stations, $S_1, \cdots, S_N$, each with one antenna, as shown in Fig.~\ref{diagram}. The stations communicate via a relay $R$ with $N$ antennas and there is no direct link between any two stations. Each time slot is divided into two subslots. The first subslot is for uplink transmissions from the stations to the relay; the second subslot is for downlink transmissions from the relay to the stations. We assume the two subslots are of equal duration. Each time slot realizes a switching permutation, as described below.

Consider one time slot. Let $\bx = \{x_1, \cdots, x_N\}^T$ be the vector representing the signals transmitted by the stations. We assume all stations use the same transmit power, normalized to one. Thus, $\mathbb{E}\{x_i^2\}=1,\ \forall\ i$. We also assume that $\mathbb{E}\{x_i\}=0,\ \forall\ i$, and that there is no cooperative coding among the stations so that $\mathbb{E}\{x_i x_j\}=0,\ \forall\ i\neq j$.
Let $\by=\{y_1, \cdots, y_N\}^T$ be the received signals at the relay, and $\bu=\{u_1, \cdots, u_N\}^T$  be the noise vector with i.i.d. noise samples following the complex Gaussian distribution, i.e., $u_n \sim \mathcal{N}_c(0, \sigma_r^2)$. Then
\begin{equation} \label{formula_uplink}
\by=\bH_u \bx +\bu,
\end{equation}
where $\bH_u$  is the uplink channel gain matrix.
The relay multiplies $\by$  by a beamforming matrix $\bG$ before relaying the signals. We impose a power constraint on the signals transmitted by the relay so that
\begin{equation} \label{formula_power}
\mathbb{E}\{\|\bG\by\|^2\} \le p.
\end{equation}
Combining (\ref{formula_uplink}) and (\ref{formula_power}), we have
%\begin{equation}
%\mathbb{E}[\bX^H \bH_u^H \bG^H \bG \bH_u \bX + \bu^H \bG^H \bG \bu ] = p.
%\end{equation}
%This gives
\begin{equation} \label{formula_tr_power}
\text{Tr}[\bH_u^H \bG^H \bG\bH_u ] + \text{Tr}[\bG^H \bG]\sigma _r^2  \le p.
\end{equation}
Let $\bH_d$  be the downlink channel matrix. Then, the received signals at the stations in vector form are
\begin{equation} \label{formula_receive}
\br = \bH_d \bG\by + \bw = \bH_d \bG\bH_u \bx + \bH_d \bG\bu  + \bw,
\end{equation}
where $\bw$ is the noise vector at the receiver, with the i.i.d. noise samples following the complex Gaussian distribution, i.e., $w_n \sim \mathcal{N}_c(0, \sigma^2)$.

\subsection{MIMO Switching}

Suppose that the purposes of $\bG$ are to realize a particular permutation represented by the permutation matrix $\bP$, and to amplify the signals coming from the stations. That is,
\begin{equation} \label{formula_channel}
\bH_d\bG\bH_u=\bA\bP,
\end{equation}
where $\bA=\diag\{a_1,\cdots,a_N\}$ is an ``amplification'' diagonal matrix.
%\begin{equation} \label{formula_receive}
%\br = \left[
%\begin{array}{*{20}c}
%a_1 x_{i_1} \\
%\vdots\\
%a_j x_{i_j}\\
%\vdots\\
%a_N x_{i_N}
%\end{array}
%\right]+ \bA\bP\bH_u^{-1}\bu+\bw,
%\end{equation}
%\begin{equation} \label{formula_receive}
%\br = \left[
%a_1 x_{i_1}, \cdots, a_j x_{i_j}, \cdots, a_N x_{i_N}
%\right]^T+ \bA\bP\bH_u^{-1}\bu+\bw,
%\end{equation}
Define $\hat \br=\bA^{-1}\br$, i.e., station $S_j$ divides its received signal by $a_j$. We can rewrite (\ref{formula_receive}) as
%\begin{equation}
%\hat \br = \left[
%\begin{array}{*{20}c}
%x_{i_1} \\
%\vdots\\
%x_{i_j}\\
%\vdots\\
%x_{i_N}
%\end{array}
%\right]+ \bP\bH_u^{-1}\bu+\bA^{-1}\bw,
%\end{equation}
\begin{equation}
\hat \br = \left[
x_{i_1}, \cdots, x_{i_j}, \cdots, x_{i_N}
\right]^T+ \bP\bH_u^{-1}\bu+\bA^{-1}\bw,
\end{equation}
where $S_{i_j}$ is the station transmitting to $S_j$  under the permutation $\bP$  (i.e., in row $j$ of $\bP$, element $i_j$ is one, and all other elements are zero).
Suppose that we require the received signal-to-noise ratio (SNR) of each station to be the same. Let $h_{u,(i,j)}^{(-1)}$  be element $(i,j)$ in $\bH_u^{-1}$. Then
\begin{equation}\label{formula_aj}
\sigma_r^2 \sum\limits_k {|h_{u,(i_j ,k)}^{( - 1)} } |^2  + \frac{{\sigma _{}^2 }}{{|a_j^{} |^2 }} = \sigma _e^2, \quad \forall\ j.
\end{equation}
%\begin{equation} \label{formula_aj}
%|a_j^{} | = \sqrt {\frac{{\sigma _{}^2 }}{{\sigma _e^2  - \sigma _r^2 \sum\limits_k {|h_{u,(i_j ,k)}^{( - 1)} } |^2 }}},\quad \forall\ j,
%\end{equation}
Note that $\sigma_e^2$  is the effective noise power for each station under unit signal power.

Substituting (\ref{formula_channel}) into (\ref{formula_tr_power}), we have
\begin{equation} \label{formula_expand_power}
q \triangleq  \sum\limits_{i,j} {|h_{d,(i,j)}^{( - 1)} } |^2 |a_j|^2  + {\rm{  }}\sigma _r^2 \sum\limits_{i,k} {|\sum\limits_j {h_{d,(i,j)}^{( - 1)} } a_j h_{u,(i_j ,k)}^{( - 1)} } |^2  \le p,
\end{equation}
where $h_{d,(i,j)}^{(-1)}$  is element $(i,j)$ in $\bH_d^{-1}$. Let $a_j=|a_j|e^{i\theta_j}$, then combining (\ref{formula_aj}) and (\ref{formula_expand_power}) gives
\begin{equation}  \label{formula_expand_power2}
q= \sum\limits_{i,j}  \frac{|{h_{d,(i,j)}^{( - 1)} } |^2{\sigma^2 }}{{\sigma _e^2  - \sigma _r^2 \sum\limits_k {|h_{u,(i_j ,k)}^{( - 1)} } |^2 }}  +
\sigma _r^2 \sum\limits_{i,k} {|\sum\limits_j  \frac{{h_{d,(i,j)}^{( - 1)} }h_{u,(i_j ,k)}^{( - 1)}\sigma e^{i\theta_j} }{\sqrt {{\sigma _e^2  - \sigma _r^2 \sum\limits_k {|h_{u,(i_j ,k)}^{( - 1)} } |^2 }}  }  } |^2  \le p,
\end{equation}

\medskip\noindent{\bf\emph{Problem Definition 1}}: Given $\bH_u, \bH_d, p, \sigma^2, \sigma_r^2$ , and a desired permutation $\bP$, solve for minimum $\sigma_e^2$ and the corresponding $\bG$.

\medskip\noindent{\bf\emph{Random-phase Algorithm}}: For a given set of $\theta_j$, $j=1,\cdots,N$, according to (\ref{formula_expand_power2}), $\lim \limits_{\sigma_e^2\rightarrow +\infty} q = 0$, and $\lim\limits_{\sigma_e^2 \rightarrow \max\limits_{i,j,u} \{\sigma _r^2 \sum\limits_k {|h_{u,(i_j ,k)}^{( - 1)} } |^2\}_+} q = +\infty$.
Furthermore, $q$ is a continuous function of $\sigma_e$. Thus, there exists a $\sigma_e$ such that $q=p$. Denote such a $\sigma_e$ by $\sigma_e(\theta_1,\cdots,\theta_N)$. The problem consists of finding
\begin{equation} \label{minsigma}
\sigma_e^{*}=\arg\min\limits_{\theta_1,\cdots,\theta_N} \sigma_e(\theta_1,\cdots,\theta_N).
\end{equation}
We note that $\sigma_e$ is a complicated nonlinear function of $\theta_j$. A time-consuming exhaustive search can be used to find the solution to (\ref{minsigma}). We use a random-phase algorithm to reduce the complexity.
%If the optimal phases of $\bA$ were known, it can be shown this problem is solvable. There are $N$ equations in (\ref{formula_aj}) and one equation in (\ref{formula_expand_power}). Then given the phases of $\bA$, the $N+1$ equations can be used to solve $a_j$  for $j=1,\cdots,N$  and $\sigma_e^2$. After that (\ref{formula_channel}) can be used to find $\bG$ from $\bA$.
We divide the interval of $[0,2\pi)$ equally into $M$ bins with the values of $0,\frac{2\pi}{M},\cdots,\frac{2(M-1)\pi}{M}$ respectively and randomly pick among them to set the the value of $\theta_j$ for each and every $j=1,\cdots,N$. After that, we compute the corresponding $\sigma_e(\theta_1,\cdots, \theta_N)$ by solving (\ref{formula_expand_power2}) with the inequality set to equality. We perform $L$ trials of these random phase assignments to obtain $L$ different values of $\sigma_e$. We choose the smallest among them as our estimate for $\sigma_e^*$. Substituting the estimated $\sigma_e^*$ into (\ref{formula_aj}) yields $|a_j|$ for all $j$; hence $\bG$. Note that the solution found is a feasible solution and is in general larger than the actual optimal $\sigma_e^*$. In Section IV, we will show that large gains can be achieved with only small $M$ and $L$. Moreover, increasing $M$ and $L$ further yields very little improvement, suggesting that the estimated $\sigma_e^*$ with small $M$ and $L$ is near optimal.

%\subsection{Comment}
%
%In multi-way relay networks, most prior works focus on two patterns of transmissions. The first is the full data exchange, in which each node needs to broadcast to all the other nodes \cite{Cui08, Gao09}. The second is pairwise unicast, in which nodes form pair, and the two nodes of a pair communicate with each other only \cite{den09, moh09}.
%In practice, however, the actual transmission patterns could be different from these two patterns. For example, in a video conference session, a subset of nodes within the network forms a multicast group, and the transmission pattern is somewhere between the two extremes above. More generally, in the same network, there could be the co-existence of broadcast sessions, multicast sessions, pairwise unicast sessions, and unidirectional unicast sessions. The MIMO switching framework here is flexible and encompasses this generality\footnote{The permutation/switch matrix $\bP$ discussed in Part B of this section can be generalized to accommodate multicast sessions. In particular, to support multicast sessions, instead of a permutation matrix, each column of the switch matrix can contain more than one ``1'' element. Each row, however, still has at most one ``1'' element.}.

\section{Fair Switching} \label{sec.STHardDec}

As has been described in the previous section, in each time slot, the stations transmit to one another according to a switch matrix. In this section, we study the fair switching scenario in which each station has an equal amount of traffic for every other station. The data from station $i$ to station $j$ could be different for different $j$, so this is not restricted to the multicast or broadcast setting. To achieve fair switching, multiple transmissions using a succession of different switch matrices over different time slots will be needed. We next discuss the set of switch matrices.

\subsection{Derangement} \label{sec.STHardDec.existing}

We assume a station does not transmit traffic to itself. A derangement is a permutation in which $i$ is not mapped to itself \cite{Has03}. While the number of distinct permutations with $N$ stations is $N!$, the number of derangements is given by the recursive formula
\begin{equation}
d_N = N\cdot d_{N-1}+(-1)^N,
\end{equation}
where $d_1=0$. For example, $d_4=9$ although the number of permutations is $4!=24$.
It can be shown that $\lim_{N\rightarrow \infty}\frac{d_N}{N!}=e^{-1}$ and the limit is approached quite quickly. Thus, the number of derangements is in general very large for large $N$. Performing optimization over this large combinatorial set of derangements in our problem is a formidable task. For example, in our fair switching problem, we want to maximize the system throughput by scheduling over a subset of derangements. It would be nice if for our problem, the optimal solution is not very sensitive to the particular selection of derangements. In Part B, we will formalize the concept of ``condensed derangement sets''.

\subsection{Condensed Derangement Set}

\medskip\noindent{\bf\emph{Definition 1}}: A set of $N-1$ derangements, $\bD_1$, $\bD_2$, $\cdots$, $\bD_{N-1}$, is said to be a \emph{condensed derangement set} if
\begin{equation} \label{formula_condensed_requirement}
\sum\limits_{n=1}^{N-1} \bD_{n}=\bJ-\bI,
\end{equation}
where $\bJ$ is a matrix with all ``1'' elements, and $\bI$ is the identity matrix.

%To find all condensed derangement sets is a combinatorial problem. We can solve it with the help of computer program. An easy algorithm to implement is we divide all the derangements into $N-1$ groups, and the matrices in one group have the same first row. Then divide the derangements in one group according to the second row, and continue until each subgroup only has one derangement. Then in order to constitute one complete derangement set, we should choose one derangement in each of the $N-1$ group. The key idea is to find $N-1$ derangements without elements 1 in the same row and column position of the matrix. Thus, we start to check the $N-1$ groups from the second row, since they are divided by the first row. It is kind of a ``smart enumeration''. There might be better search algorithms, while this one is very easy to implement.

%For $N=3$ there is only one complete derangement subset,
%\begin{equation}
%\bP_1 + \bP_2 = \bJ-\bI.
%\end{equation}
The four condensed derangement sets for $N=4$ are $\mathbb{Q}_1=\{\bP_1, \bP_5, \bP_9\}$, $\mathbb{Q}_2=\{\bP_1, \bP_6, \bP_8\}$, $\mathbb{Q}_3=\{\bP_2, \bP_4, \bP_9\}$, and $\mathbb{Q}_4=\{\bP_3, \bP_5, \bP_7\}$,  where $P_n$ are listed in TABLE I.
%\begin{equation}
%\begin{array}{l}
%\mathbb{Q}_1:\ \bP_1 + \bP_5 + \bP_9 = \bJ-\bI,\\
%\mathbb{Q}_2:\ \bP_1 + \bP_6 + \bP_8 = \bJ-\bI,\\
%\mathbb{Q}_3:\ \bP_2 + \bP_4 + \bP_9 = \bJ-\bI,\\
%\mathbb{Q}_4:\ \bP_3 + \bP_5 + \bP_7 = \bJ-\bI.
%\end{array}
%\end{equation}
There are $d_5=44$ derangements for $N=5$ and the number of condensed derangement sets is $56$.

In fair switching, we want to switch an equal amount of traffic from any station $i$ to any station $j$, $i \neq j$. This can be achieved by scheduling the derangements in the condensed derangement set in a weighted round-robin manner (as detailed in ``Approach to Problem 2'' below). Given a condensed set, the scheduling to achieve fair switching is rather simple. However, different condensed sets could potentially yield solutions of different performance. And the number of condensed derangement sets is huge for large $N$. We define a problem as follows.

\medskip\noindent{\bf\emph{Problem Definition 2}}: Suppose that we want to send equal amounts of traffic from $S_i$  to $S_j$ $\forall\ i\neq j$. Which condensed derangement sets should be used to schedule transmissions? Does it matter?

\medskip\noindent{\bf\emph{Approach to Problem 2}}: The derangements in a condensed derangement set are the building blocks for scheduling. For example, in a complete round of transmissions, we may schedule derangement $\bD_n$  for $k_n$  time slots. Then the length of the complete round transmissions will be $\sum\nolimits_{n=1}^{N-1} k_{n}$.

Consider the case of $N=4$. There are four condensed derangement sets. The question is which condensed derangement set will result in the highest throughput. We could approach the problem as follows.

Let $\mathbb{Q}_m = \{\bD_1^m, \bD_2^m, \cdots, \bD_{N-1}^m\}$ be a particular condensed derangement set. For each $\bD_n^m$ , we use random-phase algorithm above to compute the corresponding $\sigma_e^2$, denoted by $\sigma_{e,n,m}^2$. The Shannon rate is then
\begin{equation}
r_{n,m}=\log (1+\frac{1}{\sigma_{e,n,m}^2}).
\end{equation}
Because of the uniform traffic assumption, we require $k_{n,m} r_{n,m} = c$, $\forall\ n \in [1, \cdots, N-1]$,
for some $c$. That is, $c$ is the amount of traffic delivered from one station to another station in one round of transmissions. The effective throughput per station (i.e., the amount of traffic from a station to all other stations) is
\begin{equation} \label{formula_thr}
T_m =  \frac{(N-1)c}{\sum\nolimits_{n=1}^{N-1} k_{n,m}} = \frac{N-1}{\sum_{n=1}^{N-1} 1/r_{n,m}}.
\end{equation}
Numerically, we could first solve for $r_{n,m}$ $\forall\ n$. Then, we apply (\ref{formula_thr}) to find the throughput.

The question we want to answer is whether $T_m$  for different $\mathbb{Q}_m$  are significantly different.
For the case of $N=4$ and $5$, we will show simulation results indicating that the throughputs of different $\mathbb{Q}_m$ are rather close, and therefore it does not matter which $\mathbb{Q}_m$ we use.

%and assuming some simple $\bH_u, \bH_d$ , I have not been able to find significantly different $T_m$  for the four different complete permutation subsets. I did this by hand computation. For large-scale testing for different $\bH_u, \bH_d, p, \sigma^2, \sigma_r^2$ , and larger $N$, we need to the help of a computer program.

\subsection{Generalization}

As mentioned in the introduction, most prior works for multi-way relay networks focus on two patterns of transmissions. The first is pairwise unicast, in which stations form pairs, and the two stations of a pair only communicate with each other \cite{den09, moh09}. The second is the full data exchange, in which each station needs to broadcast to all the other stations \cite{Cui08, Gao09}. In practice, however, the actual transmission patterns could be different from these two patterns. For example, for video conferencing, a subset of stations within the network forms a multicast group, and the transmission pattern is somewhere between the two extremes above.

More generally, in the same network, there could be the co-existence of broadcast sessions, multicast sessions, pairwise unicast sessions, and unidirectional unicast sessions. The MIMO switching framework here is flexible and encompasses this generality.
For easy explanation, our previous discussion in Part B has an implicit assumption (focus) that each station $i$ wants to send different data to different stations $j \neq i$. If we examine the scheme carefully, this assumption is not necessary. In the scheme, a station will have chances to transmit to all other stations. In particular, a station $i$ will have chances to transmit data to two different stations $j$ and $k$ in two different derangements. If so desired, station $i$ could transmit the same data to stations $j$ and $k$ in the two derangements. This observation implies that the general traffic pattern can be realized.

For illustration, let us examine how the traffic pattern of Fig.~\ref{3node_demand} can be realized. This example is a pattern consisting of the co-existence of unicast and broadcast. As has been described, the data transmission can be realized by scheduling a condensed derangement set, which is
$\bD_1 = [\bee_3,\bee_1,\bee_2]$ and $\bD_2 = [\bee_2,\bee_3,\bee_1]$, and $\bee_n$ contains $1$ in the $n$th position and zeros elsewhere.
%If we apply the first scheme, station 1 broadcasts to station 2 and 3 in one slot. Meanwhile at least one of station 2 and 3 has to keep silent in the same slot (assume station 2 to be silent). Thus, there is no way for station 2 to transmit different data to station 1 and 3 since there are only two slots scheduled for the transmission. Then we apply the second scheme.
%\[
%\begin{array}{l}
%\bD_1 = \left[ {\begin{array}{*{20}c}
%   0 & 1 & 0  \\
%   0 & 0 & 1  \\
%   1 & 0 & 0  \\
%\end{array}} \right],\quad
%\bD_2 = \left[ {\begin{array}{*{20}c}
%   0 & 0 & 1  \\
%   1 & 0 & 0  \\
%   0 & 1 & 0  \\
%\end{array}} \right].
%\end{array}
%\]
The transmitted data of station 1, 2 and 3 are respectively $[a, b, e]^T$ for $\bD_1$ and $[a, c, d]^T$ for $\bD_2$.

\section{Simulation}
In this section, we evaluate the throughputs achieved by different condensed sets. We assume that the uplink channel $\bH_u$ and downlink channel $\bH_d$ are reciprocal, i.e., $\bH_d=\bH_u^T$, and they both follow the complex Gaussian distribution $\mathcal{N}_c(\textbf{0},\bI)$. We assume the relay has the same transmit power as all the stations, i.e., $p=1$.

\begin{figure}
\centering
\includegraphics[width=7in]{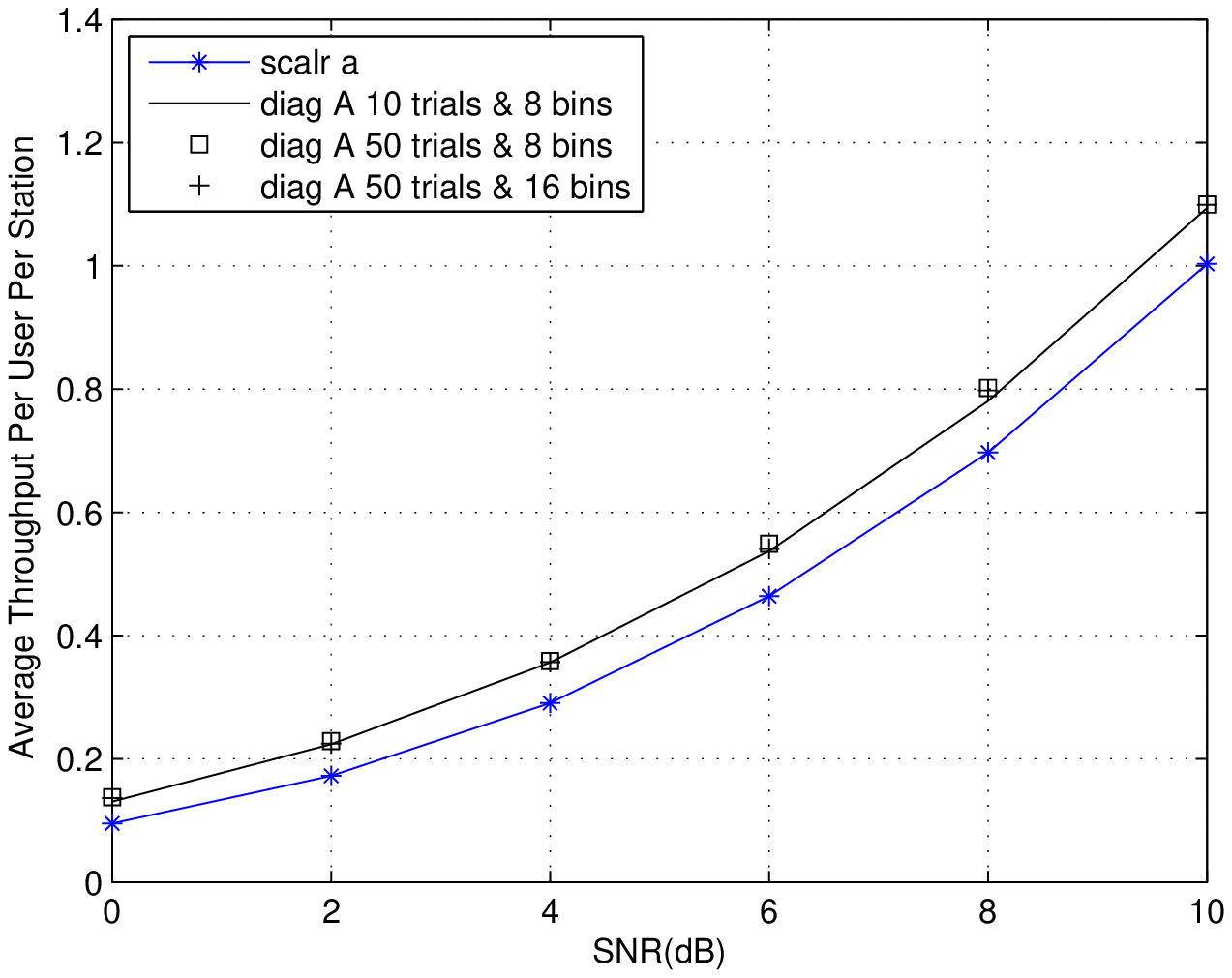}
\caption{Average throughput per station under MIMO fair switching when $N=4$. In each case only the result of one condensed derangement set is presented because the results of other derangement sets are within 2\% of the results shown here.}
\label{setsele}
\end{figure}

\begin{figure}
\centering
\includegraphics[width=7in]{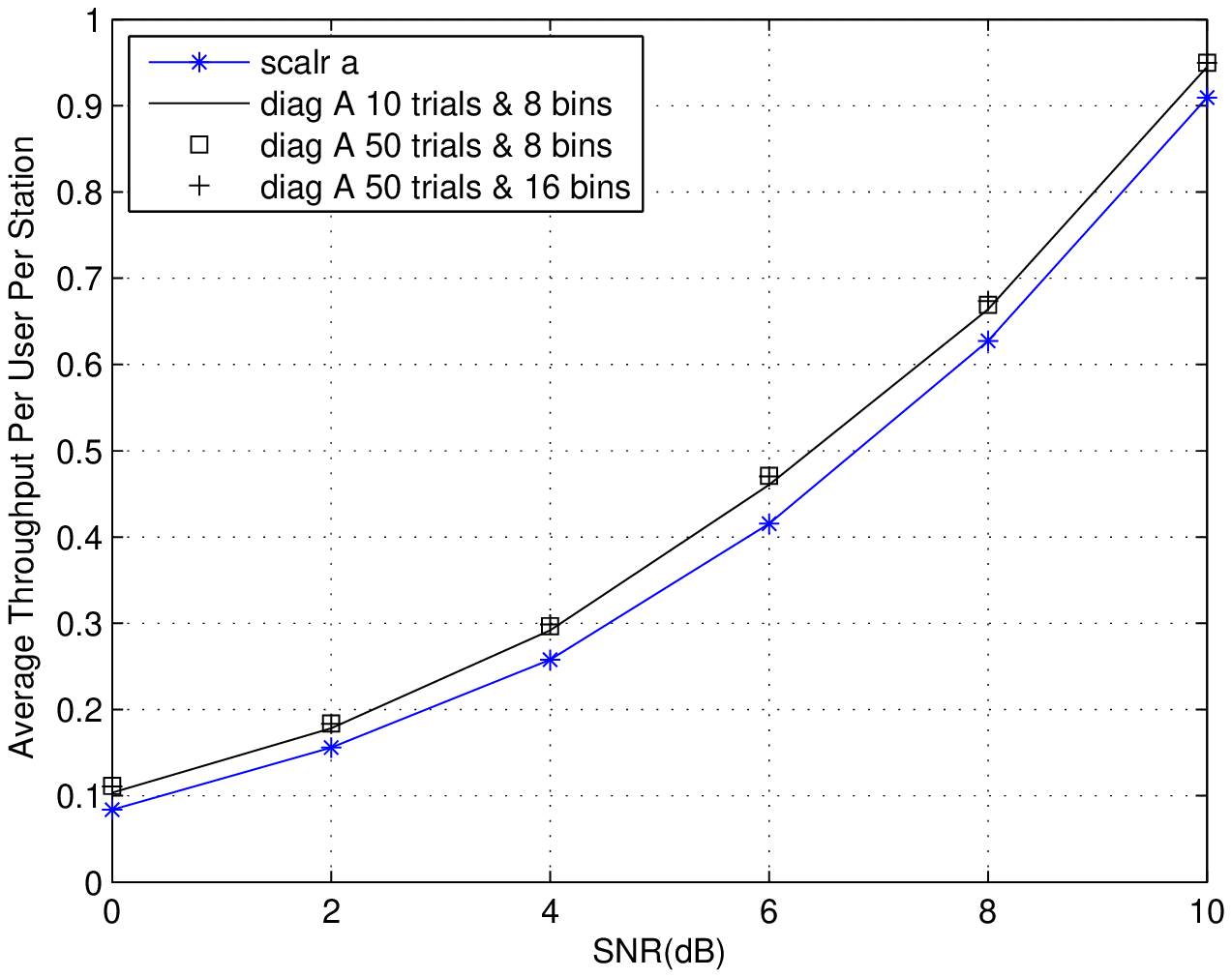}
\caption{Average throughput per station under MIMO fair switching when $N=5$. In each case only the result of one condensed derangement set is presented because the results of other derangement sets are within 1\% of the results shown here.}
\label{setsele2}
\end{figure}

We will answer the question raised in Problem Definition 2. We analyze the scenarios where $N=4$ and $N=5$. The four different condensed derangement sets of $N=4$, $\mathbb{Q}_1$, $\mathbb{Q}_2$, $\mathbb{Q}_3$ and $\mathbb{Q}_4$, are considered for fair switching.
For each channel realization, we evaluate the throughput per station $T_m$ as defined by (\ref{formula_thr}). We simulated a total of 10000 channel realizations and computed $\mathbb{E}\{T_m\}$ averaged over the channel realizations.
Recall that for random-phase algorithm, there are two associated parameters: number of trials $L$ and number of bins $M$ (see Section II). We find that for a fixed $L$ and a fixed $M$, the four condensed derangement sets yield essentially the same average throughput (within 1\% in the medium and high SNR regimes and within 2\% in the low SNR regime). Fig.~\ref{setsele} plots the throughput for one of the condensed derangement set for different $L$ and $M$. For $N=5$ there are $56$ different condensed derangement sets. As with the $N=4$ case, all the sets have roughly the same average throughput (within 1\%). Fig.~\ref{setsele2} also plots the results of one set. We also note that increasing $L$ and $M$ beyond $10$ and $8$ respectively yields little throughput gain. This implies that our heuristic yields near optimal result when $L=10$ and $M=8$.

We conjecture that different condensed derangement sets achieve roughly the same average throughput for $N$ larger than $5$ as well. A concrete proof remains an open problem. The ramification of this result, if valid, is as follows. For large $N$, the number of condensed derangement set is huge, and choosing the optimal set is a complex combinatorial problem. However, if their relative performances do not differ much, choosing any one of them in our engineering design will do, significantly simplifying the problem.

A scheme proposed in \cite{amah09} investigates a similar problem as ours. It simply uses a positive scalar weight to control the relay power consumption instead of our diagonal $\bA$. As a comparison, we also plot the throughputs of the scalar scheme in \cite{amah09}. Our scheme with diagonal $\bA$ outperforms the scalar scheme by $1$dB and $0.5$dB in Fig.~\ref{setsele} and Fig.~\ref{setsele2}. Besides the advantage in throughput, our scheme has another advantage over the scalar scheme in that our scheme guarantees fairness. That is, in our basic scheme, each station has exactly the same throughput, while the stations in the scheme in \cite{amah09} could have varying throughputs. The scalar scheme in \cite{amah09} focuses on optimizing the sum rate of all stations; the individual rates of the stations may vary widely with only one degree of freedom given by the scalar.

To sum up this section, we state the following general result:

\medskip\noindent{\bf\emph{General Result}}: In our framework of MIMO fair switching with $4$ or $5$ stations, any condensed derangement set can be used because different condensed derangement sets achieve roughly the same average throughput. We conjecture that this will be the case when the number of stations is large as well. If this conjecture holds, then the issue of condensed set selection will go away, and the complexity of the optimization problem will be greatly reduced. This conjecture remains to be proven.

%\medskip\noindent{\bf\emph{General Result 2}}: Network coding can be applied in MIMO switching to significantly improve average throughput performance. It is worth mentioning that network coding helps not only for the traditional pairwise switching pattern but also for the non-pairwise pattern. The non-pairwise pattern has not been treated in the existing literature prior to this paper.
%
%\medskip\noindent{\bf\emph{General Result 3}}: For pairwise switch without network coding, the counter-phase scheme outperforms the basic scheme with real $\bA$. Furthermore, the counter-phase scheme has low complexity because its computation does not optimize over the phases of $\bA$.

\section{Conclusions}\label{sec.Conclusion}
%In this paper we propose a framework of wireless MIMO switching for multiple stations communicating with each other, in which a multi-antenna relay controls which stations are connected to which others with beamforming. By scheduling a set of beamforming matrices (i.e. switch matrices), full connectivity among the stations can be established.

We have proposed a framework for wireless MIMO switching to facilitate communications among multiple wireless stations. A salient feature of our MIMO switching framework is that it can cater to general traffic patterns consisting of a mixture of unicast traffic, multicast traffic, and broadcast traffic flows among the stations.

There are many nuances and implementation variations arising out of our MIMO switching framework. In this paper, we have only studied the ``fair switching'' setting in which each station wants to send equal amounts of traffic to all other stations. In this setting, we aim to deliver the same amount of data from each station $i$ to each station $j \neq i$ by scheduling a set of switch matrices. In general, many sets of switch matrices could be used for such scheduling. The problem of finding the set that yields optimal throughput is a very challenging problem combinatorially. Fortunately, for number of stations $N = 4$ or $5$, our simulation results indicate that different sets of switch matrices achieve roughly the same throughput, essentially rendering the selection of the optimal set a non-issue. We conjecture this will be the case for larger $N$ as well. If this conjecture holds, then the complexity of the optimization problem can be decreased significantly as far as engineering design is concerned.

%We next moved to the study of single switch matrix and investigated the performances of different realizations for the same switch matrix (i.e., realizations using different beamforming matrices). Our general conclusion is that the use of physical-layer network coding can improve the throughput performance appreciably. In addition, we discover an interesting result for pairwise switch matrices. The computation cost of the beamforming matrix could be high in general. However, when the switch matrix is pairwise and network coding is not used, the computation of the beamforming matrix can be much simplified with a ``counter-phase'' approach.

There are many future directions going forward. For example, the beamforming matrices used in our simulation studies could be further optimized. Physical-layer network coding could be considered to improve throughput performance \cite{Zhang06physical-layernetwork}. In addition, the setting in which there are unequal amounts of traffic between stations will be interesting to explore. Also, this paper has only considered switch matrices that realize full permutations in which all stations participate in transmission and reception in each slot; it will be interesting to explore switch matrices that realize connectivities among stations that are not full permutations. Finally, future work could also explore the case where the number of antenna at the relay is not exactly $N$.

\bibliographystyle{IEEEtran}
\bibliography{MIMO_switch}

\newpage

\begin{table} \label{table}
\caption{Derangements of $N=4$.}
\[
\begin{array}{l}
\bP_1 = \qquad \qquad \quad\ \ \bP_2 = \qquad \quad \qquad\ \ \bP_3 = \\
\left[ {\begin{array}{*{20}c}
   0 & 0 & 0 & 1  \\
   0 & 0 & 1 & 0  \\
   0 & 1 & 0 & 0  \\
   1 & 0 & 0 & 0  \\
\end{array}} \right],\quad
\left[ {\begin{array}{*{20}c}
   0 & 0 & 0 & 1  \\
   0 & 0 & 1 & 0  \\
   1 & 0 & 0 & 0  \\
   0 & 1 & 0 & 0  \\
\end{array}} \right],\quad
\left[ {\begin{array}{*{20}c}
   0 & 0 & 0 & 1  \\
   1 & 0 & 0 & 0  \\
   0 & 1 & 0 & 0  \\
   0 & 0 & 1 & 0  \\
\end{array}} \right],
\end{array}
\]
\[
\begin{array}{l}
\bP_4 = \qquad \qquad \quad\ \ \bP_5 = \qquad \qquad \quad\ \ \bP_6 = \\
\left[ {\begin{array}{*{20}c}
   0 & 0 & 1 & 0  \\
   0 & 0 & 0 & 1  \\
   0 & 1 & 0 & 0  \\
   1 & 0 & 0 & 0  \\
\end{array}} \right],\quad
\left[ {\begin{array}{*{20}c}
   0 & 0 & 1 & 0  \\
   0 & 0 & 0 & 1  \\
   1 & 0 & 0 & 0  \\
   0 & 1 & 0 & 0  \\
\end{array}} \right],\quad
\left[ {\begin{array}{*{20}c}
   0 & 0 & 1 & 0  \\
   1 & 0 & 0 & 0  \\
   0 & 0 & 0 & 1  \\
   0 & 1 & 0 & 0  \\
\end{array}} \right],
\end{array}
\]
\[
\begin{array}{l}
\bP_7 = \qquad \qquad \quad\ \ \bP_8 = \qquad \qquad \quad\ \ \bP_9 = \\
\left[ {\begin{array}{*{20}c}
   0 & 1 & 0 & 0  \\
   0 & 0 & 1 & 0  \\
   0 & 0 & 0 & 1  \\
   1 & 0 & 0 & 0  \\
\end{array}} \right],\quad
\left[ {\begin{array}{*{20}c}
   0 & 1 & 0 & 0  \\
   0 & 0 & 0 & 1  \\
   1 & 0 & 0 & 0  \\
   0 & 0 & 1 & 0  \\
\end{array}} \right],\quad
\left[ {\begin{array}{*{20}c}
   0 & 1 & 0 & 0  \\
   1 & 0 & 0 & 0  \\
   0 & 0 & 0 & 1  \\
   0 & 0 & 1 & 0  \\
\end{array}} \right].
\end{array}
\]
\end{table}

\clearpage

\tikzstyle{place}=[circle,draw=blue!50,fill=blue!20,thick,
inner sep=0pt,minimum size=6mm]
\tikzstyle{transition}=[rectangle,draw=black!100,fill=none,semithick,
inner sep=0pt,minimum size=6mm]
\tikzstyle{mimoswitch}=[rectangle,draw=black!100,fill=none,semithick,
inner sep=0pt,minimum width=12mm, minimum height=26mm]
\tikzstyle{relay}=[rectangle,draw=black!100,fill=none,semithick,
inner sep=0pt,minimum width=25mm, minimum height=10mm]

\def\antenna{%
    -- +(0mm,4.0mm) -- +(2.625mm,7.5mm) -- +(-2.625mm,7.5mm) -- +(0mm,4.0mm)
}

\begin{figure}
\centering
\begin{tikzpicture}
\node (1s1) at (-1,1) [transition] {1};
\node (1s2) at ( .5,2) [transition] {2};
\node (1s3) at ( .5,0) [transition] {3};
\path[->] (1s1) edge node  [above]      {$a$} (1s2)
                edge node  [below] {$a$} (1s3);
\node (2s2) at ( 2,1) [transition] {2};
\node (2s1) at ( 3.5,2) [transition] {1};
\node (2s3) at ( 3.5,0) [transition] {3};
\path[->] (2s2) edge node [above]{$b$} (2s1)
                edge node [below]{$c$} (2s3);
\node (3s3) at ( 5,1) [transition] {3};
\node (3s1) at ( 6.5,2) [transition] {1};
\node (3s2) at ( 6.5,0) [transition] {2};
\path[->] (3s3) edge node [above] {$d$} (3s1)
                edge node [below] {$e$} (3s2);
\end{tikzpicture}
\caption{Traffic demand of a three stations example.}
\label{3node_demand}
\end{figure}
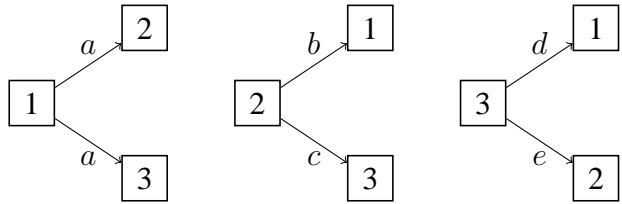

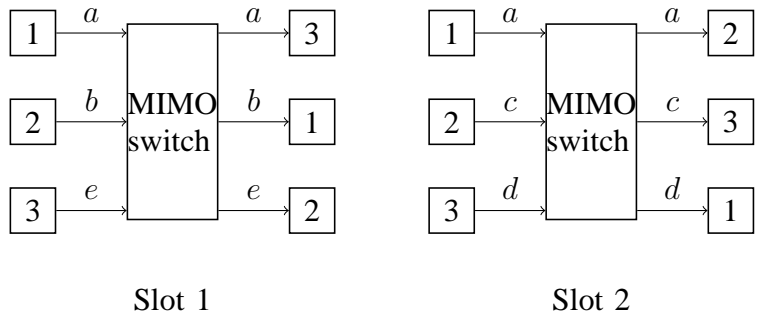
\begin{figure}
\centering
\begin{tikzpicture}[x=2.2em,y=2.8em]
\node (1s1) at (-1,2) [transition] {1};
\node (1s2) at (-1,1) [transition] {2};
\node (1s3) at (-1,0) [transition] {3};

\node (2s3) at (3,2) [transition] {3};
\node (2s1) at (3,1) [transition] {1};
\node (2s2) at (3,0) [transition] {2};

\node (m1) at (1,1) [mimoswitch] {\parbox{12mm}{MIMO \\ switch}};
\node (m2) at (7,1) [mimoswitch] {\parbox{12mm}{MIMO \\ switch}};

\node (3s1) at (5,2) [transition] {1};
\node (3s2) at (5,1) [transition] {2};
\node (3s3) at (5,0) [transition] {3};

\node (4s2) at (9,2) [transition] {2};
\node (4s3) at (9,1) [transition] {3};
\node (4s1) at (9,0) [transition] {1};

\path[->] (1s1) edge node [above]{$a$} (m1.west|-1s1);
\path[->] (m1.east|-2s3) edge node [above]{$a$} (2s3);
\path[->] (1s2) edge node [above]{$b$} (m1.west|-1s2);
\path[->] (m1.east|-2s1) edge node [above]{$b$} (2s1);
\path[->] (1s3) edge node [above]{$e$} (m1.west|-1s3);
\path[->] (m1.east|-2s2) edge node [above]{$e$} (2s2);

\path[->] (3s1) edge node [above]{$a$} (m2.west|-3s1);
\path[->] (m2.east|-4s2) edge node [above]{$a$} (4s2);
\path[->] (3s2) edge node [above]{$c$} (m2.west|-3s2);
\path[->] (m2.east|-4s3) edge node [above]{$c$} (4s3);
\path[->] (3s3) edge node [above]{$d$} (m2.west|-3s3);
\path[->] (m2.east|-4s1) edge node [above]{$d$} (4s1);

\node at (1,-1) {Slot 1};
\node at (7,-1) {Slot 2};
\end{tikzpicture}
\caption{A transmission established by two slots of unicast connectivity realizes the traffic demand in Fig.~\ref{3node_demand}.}
\label{3node_realization}
\end{figure}

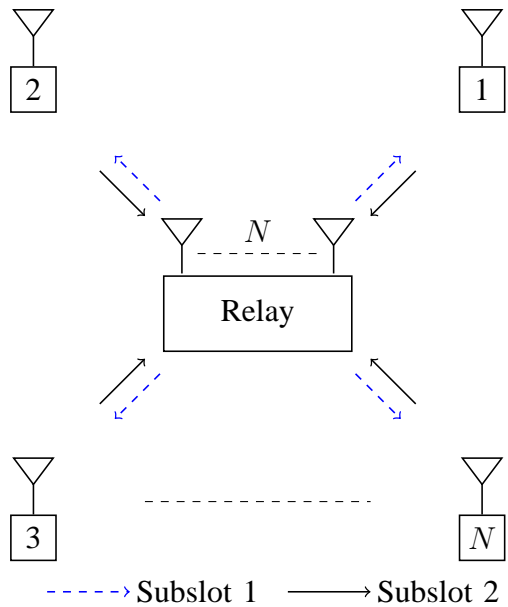
\begin{figure}
\centering
\begin{tikzpicture}[node distance=10em]
\node (relay) [relay] {Relay};
\node (1) [above right of=relay] [transition] {1};
\node (2) [above left of=relay] [transition] {2};
\node (3) [below left of=relay] [transition] {3};
\node (N) [below right of=relay] [transition] {$N$};

\draw[color=black,semithick] (1.north) \antenna;
\draw[color=black,semithick] (2.north) \antenna;
\draw[color=black,semithick] (3.north) \antenna;
\draw[color=black,semithick] (N.north) \antenna;
\path (relay.north west) to node (a) [pos=.2,inner sep=0] {} (relay.north) to node (b) [pos=.8,inner sep=0] {} (relay.north east);
\draw[color=black,semithick] (a) \antenna  (b) \antenna;

\draw [dashed] (-.8,.8) to node [above]  {$N$} (.8,.8);
\draw [dashed] (-1.5,-2.5) to node [auto] {} (1.5,-2.5);
\draw [color=blue, dashed, semithick, ->] (1.3,1.5) -- (1.9,2.1); \draw [semithick, ->] (2.1,1.9) -- (1.5,1.3);
\draw [color=blue, dashed, semithick, ->] (-1.3,1.5) -- (-1.9,2.1); \draw [semithick, ->] (-2.1,1.9) -- (-1.5,1.3);
\draw [color=blue, dashed, semithick, ->] (1.3,-0.8) -- (1.9,-1.4); \draw [semithick, ->] (2.1,-1.2) -- (1.5,-0.6);
\draw [color=blue, dashed, semithick, ->] (-1.3,-0.8) -- (-1.9,-1.4); \draw [semithick, ->] (-2.1,-1.2) -- (-1.5,-0.6);

\draw [color=blue, dashed, semithick, ->] (-2.8,-3.7) -- (-1.7,-3.7); \draw [semithick, ->] (.4,-3.7) -- (1.5,-3.7);
\node at (-0.8,-3.7) {Subslot 1};
\node at (2.4,-3.7) {Subslot 2};
\end{tikzpicture}
\caption{Wireless MIMO switching.}
\label{diagram}
\end{figure}

%\begin{figure}
%\centering
%\includegraphics[width=2.5in]{diagram}
%\caption{Wireless MIMO switching.}
%\label{diagram}
%\end{figure}

%\begin{figure}
%\centering
%\includegraphics[width=7in]{thr_cmp}
%\caption{Throughput comparison of different MIMO switching schemes for pairwise switching pattern.}
%\label{thr_cmp}
%\end{figure}
%
%
%
%\begin{figure}
%\centering
%\includegraphics[width=7in]{thr_cmp2}
%\caption{Throughput comparison of different MIMO switching schemes for non-pairwise switching pattern.}
%\label{thr_cmp2}
%\end{figure}

\end{document}

%% file: pream_bm.tex
\newtheorem{prop}{Proposition}

\newtheorem{cor}{Corollary}

\newtheorem{lm}{Lemma}

\newtheorem{thm}{Theorem}

\newcommand{\be}{\begin{eqnarray}}
\newcommand{\ee}{\end{eqnarray}}
\newcommand{\benn}{\begin{eqnarray*}}
\newcommand{\eenn}{\end{eqnarray*}}
\def\IR{\rm I \kern-0.20em R}
\newcommand{\utwi}[1]{\mbox{\boldmath $ #1$}}

\newcommand{\bthm}{\begin{thm}}
\newcommand{\ethm}{\end{thm}}

\newcommand{\bcor}{\begin{cor}}
\newcommand{\ecor}{\end{cor}}
\newcommand{\bprop}{\begin{prop}}
\newcommand{\eprop}{\end{prop}}
\newcommand{\blm}{\begin{lm}}
\newcommand{\elm}{\end{lm}}
\newcommand{\beq}{\begin{equation}}
\newcommand{\eeq}{\end{equation}}
\newcommand{\ber}{\begin{eqnarray}}
\newcommand{\eer}{\end{eqnarray}}

\newcommand{\bproof}{\begin{proof}}
\newcommand{\eproof}{\end{proof}}

\newcommand{\diag}{\mathop{\mbox{\rm diag}}}

\newcommand{\bit}{\begin{itemize}}
\newcommand{\eit}{\end{itemize}}
\newcommand{\ben}{\begin{enumerate}}
\newcommand{\een}{\end{enumerate}}
\newcommand{\bdesc}{\begin{description}}
\newcommand{\edesc}{\end{description}}
\newcommand{\beqarrn}{\begin{eqnarray*}}
\newcommand{\eeqarrn}{\end{eqnarray*}}
\newcommand{\bproofof}{\begin{proofof}}
\newcommand{\eproofof}{\end{proofof}}
\newenvironment{rem}{\begin{trivlist}\item[]{\bf
Remark:}\hspace{4mm}}{\end{trivlist}}
\newcommand{\brem}{\begin{rem}}
\newcommand{\erem}{\end{rem}}
\newenvironment{rems}{\begin{trivlist}\item[]{\bf
Remarks}\begin{itemize}}{\end{itemize}\end{trivlist}}
\newcommand{\brems}{\begin{rems}}
\newcommand{\erems}{\end{rems}}
\newtheorem{fact}{Fact}
\newcommand{\bfact}{\begin{fact}}
\newcommand{\efact}{\end{fact}}
\newtheorem{examp}{Example}
\newcommand{\bexamp}{\begin{examp}\rm}
\newcommand{\eexamp}{\end{examp}}
\newtheorem{defn}{Definition}
\newcommand{\bdefn}{\begin{defn}\rm}
\newcommand{\edefn}{\end{defn}}

\newtheorem{alg}{Algorithm}
\newcommand{\balg}{\begin{alg}}
\newcommand{\ealg}{\end{alg}}

\newtheorem{prob}{Problem}
\newcommand{\bprob}{\begin{prob}}
\newcommand{\eprob}{\end{prob}}

\newcommand{\bvtm}{\begin{verbatim}}
\newcommand{\bfig}{\begin{figure}}
\newcommand{\efig}{\end{figure}}
\newcommand{\bcen}{\begin{center}}
\newcommand{\ecen}{\end{center}}

\long\def\comment#1{}

\def \n2{{N_0 \over 2}}

\def \h5{\hspace{0.5in}}

\newcommand{\bee}{{\utwi{e}}}

\newcommand{\br}{{\utwi{r}}}

\newcommand{\bu}{{\utwi{u}}}

\newcommand{\bw}{{\utwi{w}}}
\newcommand{\bx}{{\utwi{x}}}
\newcommand{\by}{{\utwi{y}}}

\newcommand{\bA}{{\utwi{A}}}

\newcommand{\bD}{{\utwi{D}}}

\newcommand{\bG}{{\utwi{G}}}
\newcommand{\bH}{{\utwi{H}}}
\newcommand{\bI}{{\utwi{I}}}
\newcommand{\bJ}{{\utwi{J}}}

\newcommand{\bP}{{\utwi{P}}}

%% file: Letter_MIMO_Switch_04152.bbl
\begin{thebibliography}{1}
\providecommand{\url}[1]{#1}
\csname url@samestyle\endcsname
\providecommand{\newblock}{\relax}
\providecommand{\bibinfo}[2]{#2}
\providecommand{\BIBentrySTDinterwordspacing}{\spaceskip=0pt\relax}
\providecommand{\BIBentryALTinterwordstretchfactor}{4}
\providecommand{\BIBentryALTinterwordspacing}{\spaceskip=\fontdimen2\font plus
\BIBentryALTinterwordstretchfactor\fontdimen3\font minus
  \fontdimen4\font\relax}
\providecommand{\BIBforeignlanguage}[2]{{%
\expandafter\ifx\csname l@#1\endcsname\relax
\typeout{** WARNING: IEEEtran.bst: No hyphenation pattern has been}%
\typeout{** loaded for the language `#1'. Using the pattern for}%
\typeout{** the default language instead.}%
\else
\language=\csname l@#1\endcsname
\fi
#2}}
\providecommand{\BIBdecl}{\relax}
\BIBdecl

\bibitem{cov79}
T.~Cover and A.~Gamal, ``Capacity theorems for the relay channel,''
  \emph{Information Theory, IEEE Transactions on}, vol.~25, no.~5, pp.
  572--584, 1979.

\bibitem{den09}
D.~Gunduz, A.~Yener, A.~Goldsmith, and H.~Poor, ``The multi-way relay
  channel,'' in \emph{Information Theory, 2009. ISIT 2009. IEEE International
  Symposium on}, 2009, pp. 339--343.

\bibitem{moh09}
Y.~Mohasseb, H.~Ghozlan, G.~Kramer, and H.~El~Gamal, ``The {MIMO} wireless
  switch: Relaying can increase the multiplexing gain,'' in \emph{Information
  Theory, 2009. ISIT 2009. IEEE International Symposium on}, 2009.

\bibitem{Cui08}
T.~Cui, T.~Ho, and J.~Kliewer, ``Space-time communication protocols for n-way
  relay networks,'' in \emph{IEEE Global Telecommunications Conference, 2008.},
  2008.

\bibitem{Gao09}
F.~Gao, T.~Cui, B.~Jiang, and X.~Gao, ``On communication protocol and
  beamforming design for amplify-and-forward n-way relay networks,'' in
  \emph{Computational Advances in Multi-Sensor Adaptive Processing (CAMSAP),
  3rd IEEE International Workshop on}, 2009.

\bibitem{Has03}
M.~Hassani, ``Derangements and applications,'' \emph{Journal of Integer
  Sequences}, vol.~6, no.~1, 2003.

\bibitem{amah09}
A.~U.~T. Amah and A.~Klein, ``Non-regenerative multi-way relaying with linear
  beamforming,'' in \emph{Proc. 20th IEEE International Symposium on Personal,
  Indoor and Mobile Radio Communications Symposium}, 2009.

\bibitem{Zhang06physical-layernetwork}
S.~Zhang, S.~C. Liew, and P.~P. Lam, ``Physical-layer network coding,'' in
  \emph{ACM Mobicom}, 2006.

\end{thebibliography}
